\documentclass[aps,prf,reprint,groupedaddress,nofootinbib,longbibliography,showpacs]{revtex4-1}
\usepackage{graphicx}
\usepackage{upgreek}
\usepackage{amsmath}
\usepackage{color}
\begin{document}

\title{\textcolor{black}{Velocity fluctuations} and boundary layer structure in a rough Rayleigh-B\'enard cell filled with water}


\author{Olivier Liot}
 \altaffiliation[Presently at ]{LAAS/CNRS, 7 avenue du Colonel Roche, 31400 Toulouse, France}
\author{Quentin Ehlinger}%
\altaffiliation[Presently at ]{Univ Lyon, Universit\'e Claude Bernard Lyon 1, CNRS, Institut Lumi\`ere Mati\`ere, Campus LyonTech- La Doua, B\^atiment Kastler, 10 rue Ada Byron, 69622 Villeurbanne CEDEX, France, EU}
\author{\'El\'eonore Rusaou\"en}
\altaffiliation[Presently at ]{LEGI, Domaine Universitaire, CS 40700, 38058 Grenoble Cedex 9, France, EU}
\author{Thibaut Coudarchet}
\author{Julien Salort}
\author{Francesca Chill\`a}
\email[]{francesca.chilla@ens-lyon.fr}
\affiliation{Univ Lyon, ENS de Lyon, Univ Claude Bernard, CNRS, Laboratoire de Physique, F-69342 Lyon Cedex 7, France}
%


\date{\today}

\begin{abstract}
We report Particle Image Velocimetry of the Large Scale Circulation and the viscous boundary layer in turbulent thermal convection. We use two parallelepipedic Rayleigh-B\'enard cells with a top smooth plate. The first one has a rough bottom plate and the second one has a smooth one so we compare the rough-smooth and the smooth-smooth configurations. The dimensions of the cell allow to consider a bi-dimensional mean flow. Lots of previous heat flux measurements have shown a Nusselt--Rayleigh regime transition corresponding to an increase of the heat flux in presence of roughness which is higher than the surface increase. Our velocity measurements show that if the mean velocity field is not clearly affected by the roughness, the velocity fluctuations rise dramatically. It is accompanied by a change of the longitudinal velocity structure functions scaling. Moreover, we show that the boundary layer becomes turbulent close to roughness, as it was observed recently in the air [Liot \emph{et al., JFM}, vol. 786, pp. 275-293]. Finally we discuss the link  between the change of the boundary layer structure and the ones observed on the \textcolor{black}{velocity fluctuations}.
\end{abstract}

\pacs{
      {47.55.pb Thermal convection}; {44.25.+f Natural convection}; {47.27.-i Turbulent flows}
     } 

\maketitle
\section{Introduction}
Buoyancy fluctuations are the engine of various flows. Atmospheric circulation or earth's mantle motions are governed by thermal convection. Lots of industrial applications (cooling of a nuclear plant for example) also use this kind of flow. Because thermal convection is often turbulent, it is a very efficient way for carrying heat. But even if this flow is very accessible and has been studied for a long time \cite{rayleigh1916,kraichnan1962}, lots of properties and mechanisms in play in turbulent thermal convection still need to be understood. In the laboratory, we have chosen to model thermal convection flows with the Rayleigh-B\'enard configuration: a horizontal layer of fluid confined between a cooling plate above and a heating plate below. The Rayleigh number measures the forcing due to buoyancy effects compared to dissipative ones:


\begin{equation}
Ra = \frac{g\alpha \Delta T H^3}{\nu \kappa},
\end{equation}

\noindent where $H$ is the height of the cell, $g$ is the acceleration due to gravity, $\alpha$ is the constant pressure thermal expansion coefficient of the fluid, $\nu$ its kinematic viscosity, $\kappa$ its thermal diffusivity and $\Delta T=T_h - T_c$ is the difference of temperature between the heating and cooling plates. The Prandtl number compares viscosity to thermal diffusivity:

\begin{equation}
Pr=\frac{\nu}{\kappa}. 
\end{equation}

\noindent A last standard parameter is the aspect ratio $\Gamma$. It is the ratio between the characteristic transverse length of the cell and its height $H$. These numbers represent the control parameters of thermal convection. We represent the response of the system by the dimensionless heat flux, the Nusselt number:

\begin{equation}
Nu = \frac{QH}{\lambda \Delta T},
\end{equation} 

\noindent where $Q$ is the global heat flux, and $\lambda$ is the thermal conductivity of the fluid. 

In turbulent Rayleigh-B\'enard convection, the mixing makes the bulk temperature almost homogeneous. Temperature gradients are confined close to the plates in thermal boundary layers. Their thickness can be computed by:

\begin{equation}
\delta_{th}=\frac{H}{2Nu}.
\label{eq:deltaT}
\end{equation}

\noindent Viscous boundary layers also develop along the plates. Thermal transfer is due to interactions between the bulk and these boundary layers. Particularly, plumes are slices of them which detach and go towards the opposite plate. They play a crucial role in thermal transfer \cite{grossmann2004} and understanding their statistics, structure and coherence is still a challenge \cite{chilla2012}. 


Some progress w\textcolor{black}{as} made in the understanding of relation between response and control parameters. Particularly, lots of efforts have been concentrated in the study of relation between thermal forcing and heat flux, $Nu\propto Ra^\gamma$  \cite{shraiman1990,grossmann2000,ahlers2009b,stevens2013}. Nevertheless, alternative methods are necessary to have a larger scope about turbulent Rayleigh-B\'enard convection \cite{chilla2012}.


One of them, used in the present paper, consists in a destabilization of the boundary layers with controlled roughness. The Hong-Kong group used pyramid-shaped roughness on both plates \cite{shen1996,du2000,qiu2005}. They observed a heat flux enhancement up to 76\% compared to a smooth case which is higher than the surface increase, even if there is not always a change for the scaling exponent $\gamma$. This enhancement is attributed to an increase of the plumes emission by the top of roughness. This result can be extended to cells where only one plate is rough \cite{wei2014}. Groove-shaped roughness were used in Grenoble \cite{roche2001}, and the scaling exponent reached $\gamma=1/2$. Numerical simulations for the same geometry showed an increase of $\gamma$ too \cite{stringano2006}. Increase of the scaling exponent has been also observed by Ciliberto \& Laroche \cite{ciliberto1999} with randomly distributed glass spheres on the bottom plate. \textcolor{black}{All these observations put forward that the heat flux enhancement relatively to a smooth configuration starts from a transitional Rayleigh number $Ra_t$.} It is now admitted that the $Nu$--$Ra$ regime transition appears when the thermal boundary layer thickness becomes similar to the roughness height $h_0$ \cite{tisserand2011}. The corresponding transition Nusselt number is:

\begin{equation}
Nu_t=\frac{H}{2h_0}.
\label{eq:transition}
\end{equation}

In Lyon, several experiments have been performed with square-studs roughness on the bottom plate, both in a cylindrical cell \cite{tisserand2011} and a parallelepipedic one \cite{salort2014} filled with water. Similar results about the heat flux increase have been observed. Moreover, very close to roughness, temperature fluctuations study led us to a phenomenological model based on a destabilization of the boundary layer. This model is in good agreement with global heat flux measurements. This destabilization was confirmed by velocity measurements inside the viscous boundary layer close to roughness in a proportional cell filled with air. These experiments in the \emph{Barrel of Ilmenau} showed that the viscous boundary layer transits to a turbulent state above roughness \cite{liot2016}.

Box-shaped roughness have been studied analytically \cite{shishkina2011} and numerically \cite{wagner2015}. It consists in four elements on the bottom plate whose height is much larger than the thermal boundary layer thickness. It is different from previous presented studies where roughness and thermal boundary layer have a similar size. An increase in $\gamma$ was observed then a saturation of the supplementary heat flux rise when the zones between roughness elements are totally washed out by the fluid. 

In this paper we present Particle Image Velocimetry (PIV) measurements performed in the whole cell and close to roughness. We used the same cell as Salort \emph{et al.} \cite{salort2014}. The bottom plate is rough whereas the top one is smooth. A similar cell is used with both smooth plates for comparison. Whereas no effect on the mean velocity field is clearly visible, a large increase of the velocity fluctuations is observed with presence of roughness, probably related to an increase of plumes emission and intensity. This is accompanied by a change of the velocity structure functions shape. Moreover, \textcolor{black}{hints of} logarithmic velocity profiles are put forward above roughness in the same way as a previous study in the air \cite{liot2016}, which is a new evidence of the transition to turbulence of the viscous boundary layer. \textcolor{black}{We remind that logarithmic temperature profiles have already been observed close to smooth plates} \cite{ahlers2012}. \textcolor{black}{They can appear without logarithmic velocity profiles}.

\section{Experimental setup and PIV acquisition}
\textcolor{black}{The first part of this paper presents the experimental setup and the velocity acquisition method.}

\subsection{Convection cells}

\begin{table}
\begin{center}
\begin{tabular}{ccccc}
\hline\noalign{\smallskip}
Cell & $\Delta T$ & $T_m=(T_h+T_b)/2$ &  $Ra$ & $Pr$ \\
\noalign{\smallskip}\hline\noalign{\smallskip}
$\mathcal{RS}$ & 25.9$^\mathrm{o}$C & 40.0$^\mathrm{o}$C &  $7.0\times 10^{10}$ & 4.4\\
$\mathcal{SS}$ & 25.7$^\mathrm{o}$C & 40.0$^\mathrm{o}$C &  $6.9\times 10^{10}$ & 4.4\\
\noalign{\smallskip}\hline
\end{tabular}
\end{center}
\caption{Parameters used for the global velocity measurements in the $\mathcal{RS}$ (rough-smooth) and $\mathcal{SS}$ (smooth-smooth) cells.}
\label{table:tableau}
\end{table}

\begin{table}
\begin{center}
\begin{tabular}{cccc}
\hline\noalign{\smallskip}
 $\Delta T$ & $T_m=(T_h+T_b)/2$ &  $Ra$ & $Pr$ \\
\noalign{\smallskip}\hline\noalign{\smallskip}
 14.8$^\mathrm{o}$C & 40.0$^\mathrm{o}$C &  $4.0\times 10^{10}$ & 4.4\\
\noalign{\smallskip}\hline
\end{tabular}
\end{center}
\caption{Experimental parameters used for the velocity measurements close to roughness.}
\label{table:tableau2}
\end{table}

\begin{figure}
\begin{center}
\includegraphics[width=8cm]{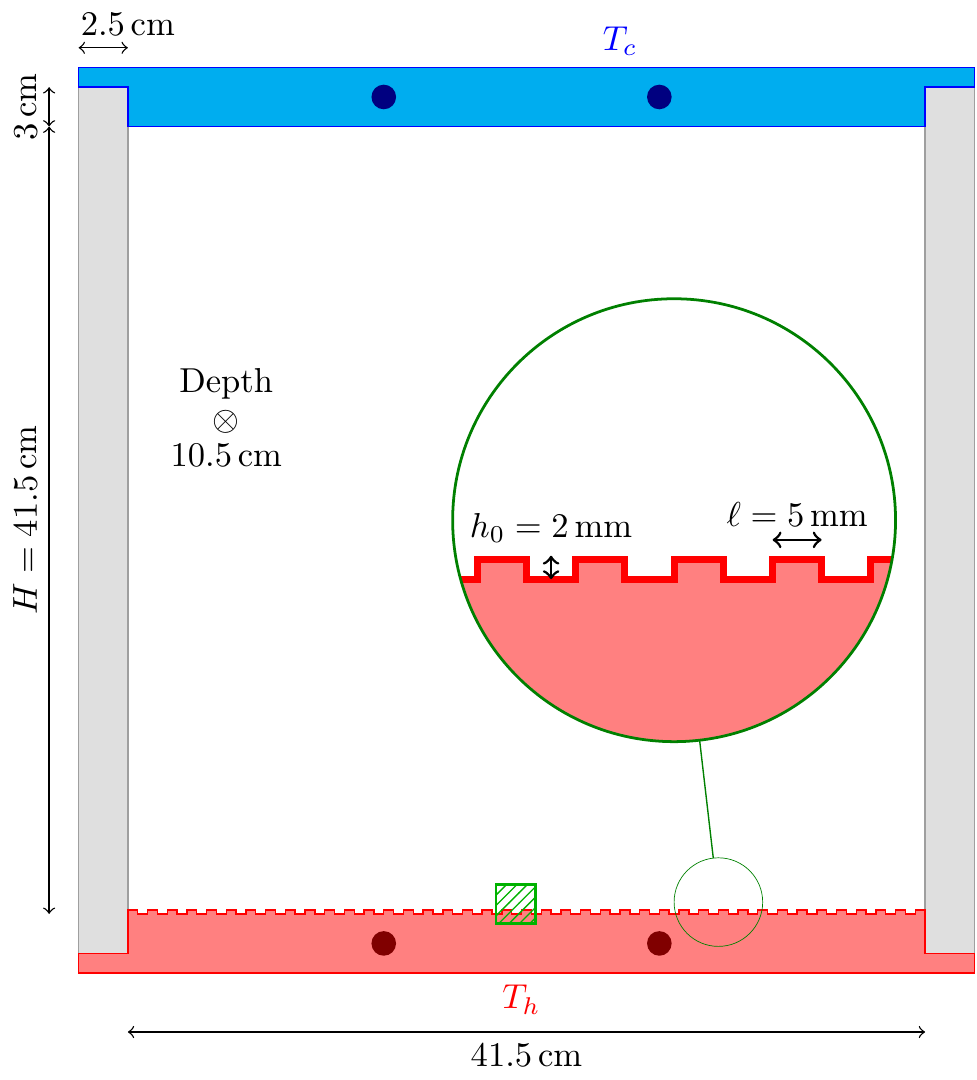}
   \caption{Sketch of the $\mathcal{RS}$ convection cell. The four dark dots in the plates show the location of the PT100 temperature sensors. The zoom shows the roughness dimensions. PIV measurements of the viscous boundary layer are performed in the green hatched area.}
   \label{cell}
   \end{center}
   \end{figure}

\begin{figure}
\begin{center}
\includegraphics[width=4.5cm]{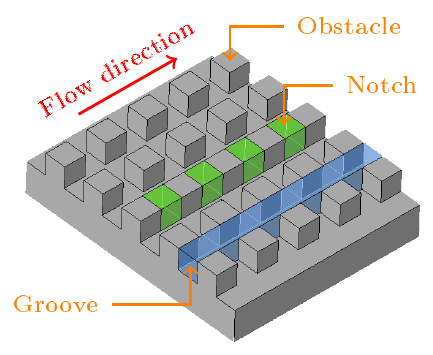}
   \caption{Detail of roughness with the three different locations according to the flow direction.}
   \end{center}
   \label{rugosites}
   \end{figure}

We use a parallelepipedic convection cell of 10.5\,cm-thick 41.5$\times$41.5\,cm$^2$ with 2.5\,cm-thick PMMA walls (see sketch figure \ref{cell}). The top plate consists in a 4\,cm-thick copper plate coated with a thin layer of nickel. The bottom plate is an aluminium alloy (5083) anodized in black. It is Joule-heated while the top plate is cooled with a temperature regulated water circulation. Plate temperatures are measured by PT100 temperature sensors. On the bottom plate, periodic roughness are machined directly in the plate. It consists in an array of 0.2\,cm-high, 0.5$\times$0.5\,cm$^2$ square obstacles (zoom figure \ref{cell}). Because of the cell dimensions, we can assume that the mean flow is quasi bi-dimensional. Thus, according to the flow direction, we can distinguish three positions close to roughness: above an obstacle, inside a notch where the fluid is "protected" from the mean flow and in the groove between obstacle rows (figure \ref{rugosites}). We call this cell "rough-smooth" (abbreviated $\mathcal{RS}$). Moreover, we use a very similar cell with a smooth bottom plate to perform reference global velocity measurements. The only difference is that the bottom plate is in copper anodized with a thin layer of nickel. This cell is named "smooth-smooth" (abbreviated $\mathcal{SS}$).

   
   The cells are filled with deionized water. The main experimental parameters are grouped in table \ref{table:tableau}. The regime transition \textcolor{black}{(i.e. when the heat flux in the $\mathcal{RS}$ configuration becomes higher than in the $\mathcal{SS}$ one)} occurs when the thermal boundary layer reaches $h_0$. We have $Nu_t=H/2h_0$ according to equation \ref{eq:transition}. If we suppose in this cell the relation $Nu=0.06\,Ra^{1/3}$ \cite{salort2014}, the Nusselt-Rayleigh regime transition occurs for $Ra_t=4.1\times 10^9$. Consequently, we work at a Rayleigh number far after the transition. Concerning the velocity measurements close to roughness, table \ref{table:tableau2} sums up the experimental conditions. Unfortunately in this cell we are not able to reach Rayleigh numbers \textcolor{black}{below} the transition \textcolor{black}{threshold} while allowing visualization and stable flow. 

%
%


\subsection{PIV acquisition}

PIV acquisitions are performed using a 1.2 W, Nd:YVO$_4$ laser. With a cylindrical lens we build a vertical laser sheet which enters in the cell from the observer's left hand side. We seed the flow with \emph{Sphericel 110P8} glass beads of 1.10$\pm$0.05 in density and 12\,$\upmu$m average diameter.

 For global velocity field acquisitions we use a digital camera \emph{Stingray F-125B}. We perform twelve-hour acquisitions with one picture pair every ten seconds (4320 picture pairs). Picture on the same pair are separated of fifty milliseconds. For analysis we use the free software CIVx \cite{fincham2000}. For the $\mathcal{RS}$ plate, we use a first pass of picture pair cross-correlation with 64$\times$64 pixels$^2$ boxes with 50$\%$ overlap. Search boxes are one and a half larger. Then other passes are used with smaller boxes. For the $\mathcal{SS}$ cell, we use the same method but with 30$\times$30 pixels$^2$ first pass boxes size. \textcolor{black}{The resulting resolution gets down to about 3\,mm in the $\mathcal{RS}$ cell and 6\,mm in the $\mathcal{SS}$ one.} For measurements close to roughness, a faster acquisition process is necessary to have a sufficient space and time resolution for PIV treatment. We use a \emph{IOI Flare 2M360CL} digital camera. Pictures are captured continuously at frequencies from 200 to 340 frames per second. \textcolor{black}{The resolution gets down to about 250\,$\mu$m.} On one hand acquisitions are performed in a groove and on the other hand above obstacles and in notches simultaneously. All of these locations are chosen at the center of the cell (see figure \ref{cell}).



\section{Study of the Large Scale Circulation}

The \textcolor{black}{second} part of this paper consists in observing global velocity fields, velocity fluctuations and velocity longitudinal structure functions in the whole cell. We compare the $\mathcal{RS}$ and $\mathcal{SS}$ cells.

\subsection{Mean velocity fields}

\begin{figure*}
\begin{center}
\includegraphics[width=16cm]{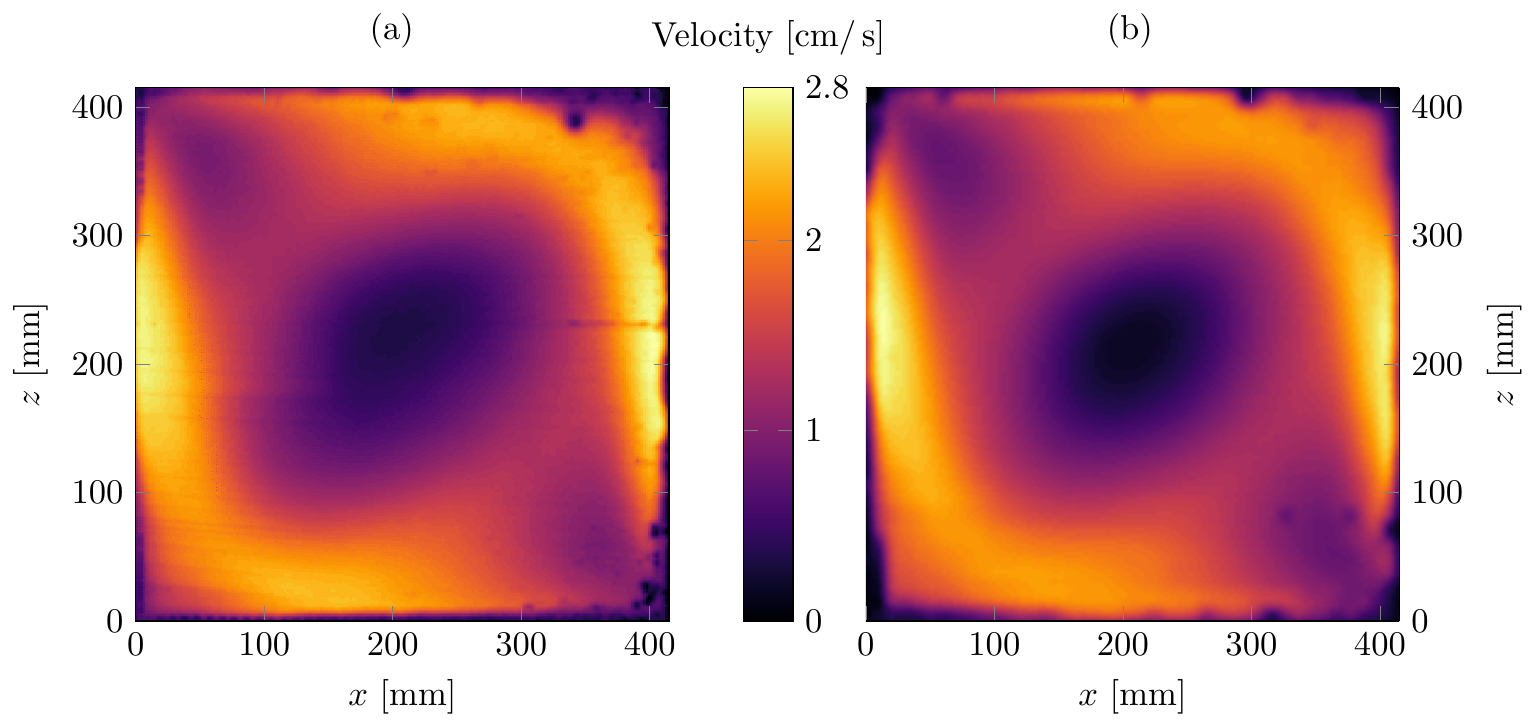}
\end{center}
\caption{Velocity magnitude fields in (a) $\mathcal{RS}$ cell and (b) $\mathcal{SS}$ cell. $Ra=7.0\times10^{10}$ and $Ra=6.9\times10^{10}$ respectively.}
\label{mod_V_rug_lisse}
\end{figure*}

First of all, we plot the mean velocity magnitude map. Figure \ref{mod_V_rug_lisse} compares the $\mathcal{RS}$ case to the $\mathcal{SS}$ one. We proceed at close Rayleigh number: $Ra=7.0\times10^{10}$ and $Ra=6.9\times10^{10}$ respectively which allows a direct comparison of the measurements. In both cases we choose acquisitions where the LSC occurs counter-clockwise. Hot and cold jets spread along the right and left sidewall respectively. Corresponding velocities are around 2.2\,cm/s whereas in the center part of the flows, the velocity magnitude is quite slower ($\sim 0.5$\,cm/s). The $\mathcal{SS}$ velocity field is very similar to that obtained by Xia \emph{et al.} \cite{xia2003} in a proportional cell filled with water for $Ra=3.5\times10^{10}$. We observe that the LSC structure is very similar for both $\mathcal{RS}$ and $\mathcal{SS}$ cells. Nevertheless, the velocity in the hot jet seems a little bit larger for the $\mathcal{RS}$ case. The mean velocity in the $\mathcal{SS}$ cell is $1.36\pm0.01$\,cm/s against $1.43\pm0.01$\,cm/s in the $\mathcal{RS}$ one which corresponds to a 5\% increase. If it is not a large difference, there is a possible effect of roughness because the Rayleigh number for the $\mathcal{RS}$ case is only 1.5\% higher than for the $\mathcal{SS}$ one which corresponds to a negligible Reynolds number increase of about 1\%. This observation lets us think that there is a small effect of roughness on the mean velocity field. But this small velocity difference may fall within the experimental uncertainties due to parallax, calibration of laser sheet orientation.

\subsection{Velocity fluctuations}

\begin{figure*}
\begin{center}
\includegraphics[width=16cm]{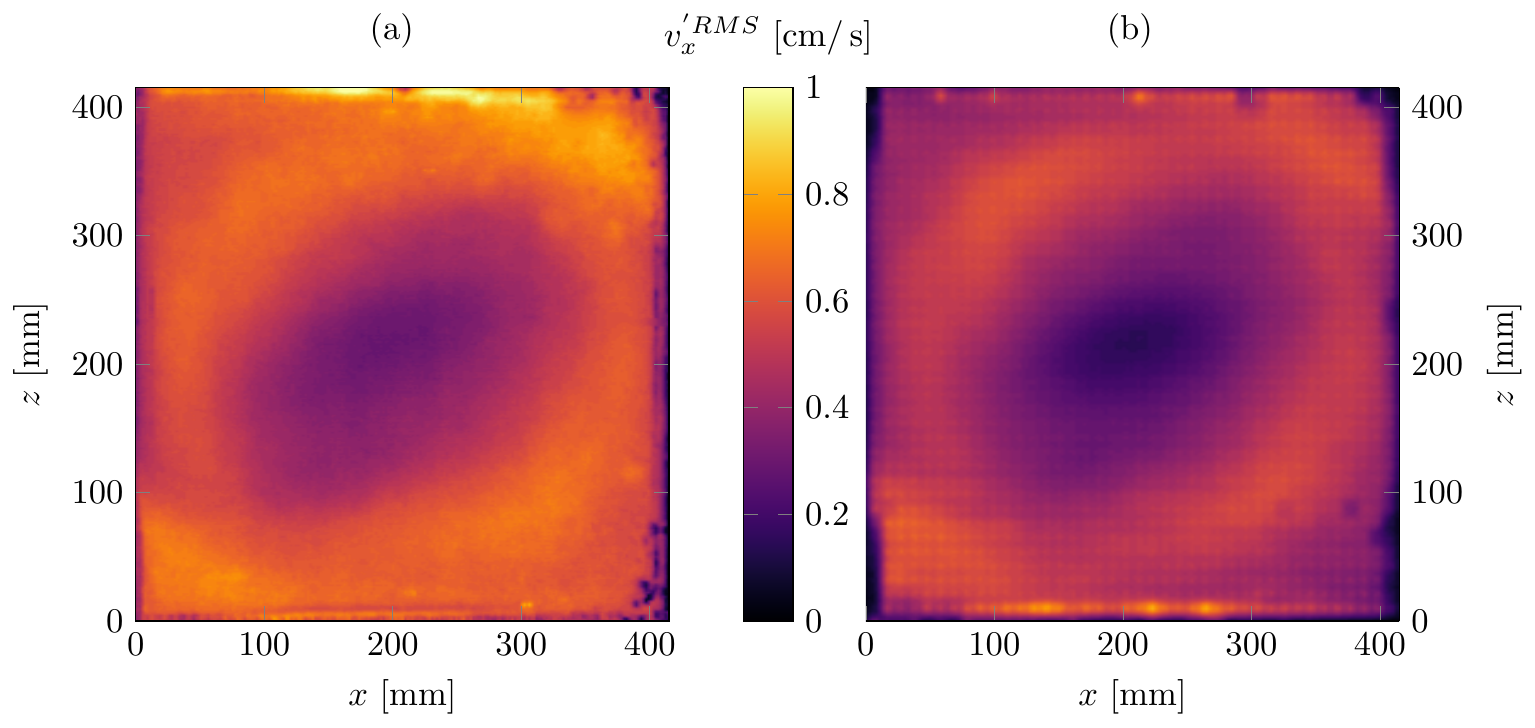}
\end{center}
\caption{Horizontal velocity fluctuations RMS field in the (a) $\mathcal{RS}$ and (b) $\mathcal{SS}$ cells. Counter-clockwise LSC; $Ra=7.0\times10^{10}$ and $Ra=6.9\times10^{10}$ respectively.}
\label{RMSvx}
\end{figure*}

\begin{figure*}
\begin{center}
\includegraphics[width=16cm]{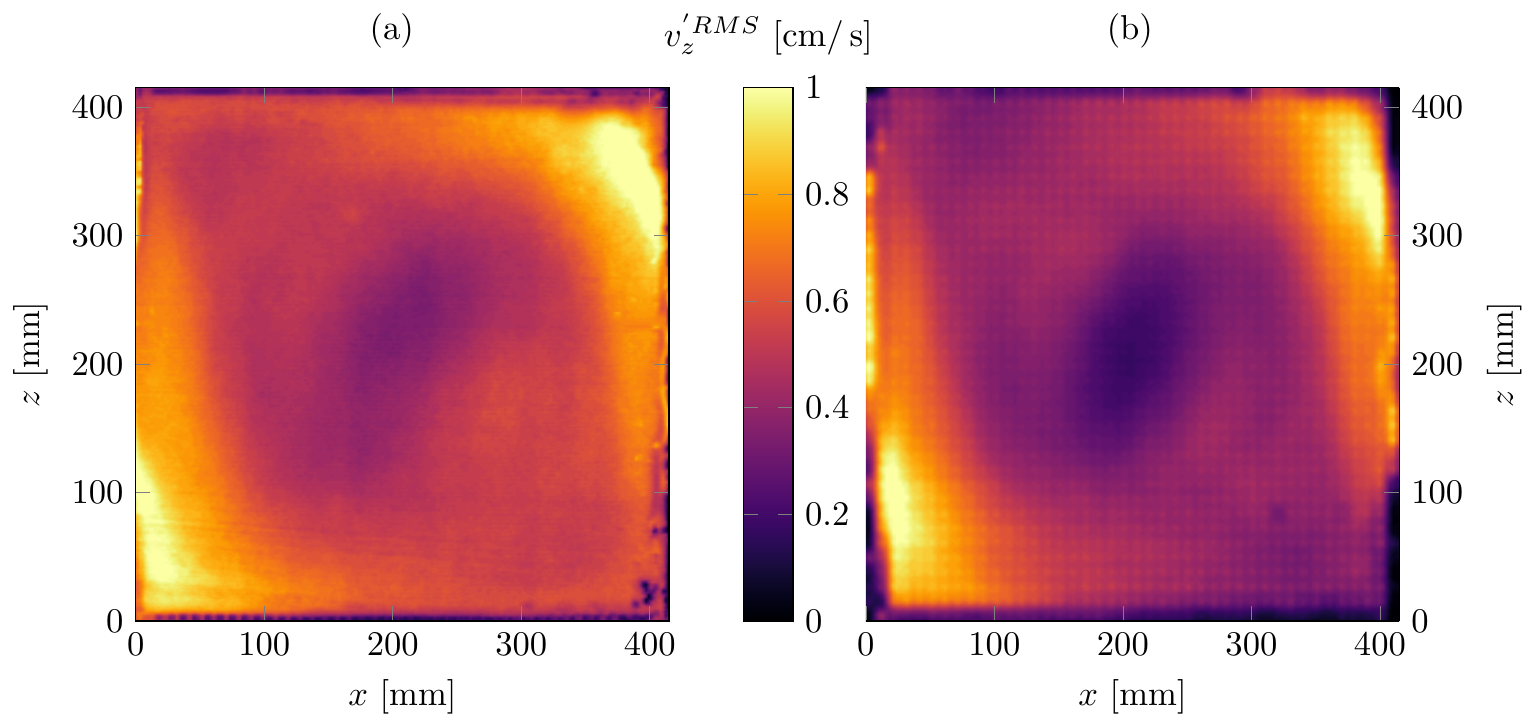}
\end{center}
\caption{Vertical velocity fluctuations RMS field in (a) $\mathcal{RS}$ and (b) $\mathcal{SS}$ cells. Counter-clockwise LSC; $Ra=7.0\times10^{10}$ and $Ra=6.9\times10^{10}$ respectively.}
\label{RMSvy}
\end{figure*}

Because we do not see a clear influence of roughness on the mean velocity fields, we try to observe it on the velocity fluctuations. We compute the velocity fluctuations root mean square (RMS) for the horizontal ($v'_x$) and vertical ($v'_z$) velocity fluctuations as: 

\begin{equation}
v^{'RMS}_{i}(x,z)=\sqrt{\langle\left( v_{i}(x,z,t)-\langle v_{i}(x,z,t)\rangle_t\right)^2\rangle_t},
\end{equation} 

\noindent where $i=x,z$. With this definition, we report in figures \ref{RMSvx} and \ref{RMSvy} the RMS values of the horizontal and vertical components respectively. The column (a) is for the $\mathcal{RS}$ configuration, the column (b) for the $\mathcal{SS}$ one.

\begin{figure}
\begin{center}
\includegraphics[width=8cm]{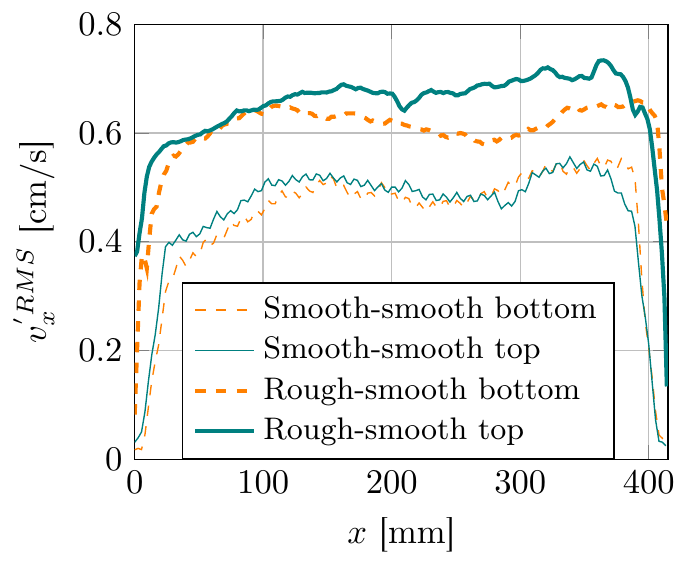}
\caption{Profiles of average fields of horizontal velocity RMS. Profiles of "bottom" region are horizontally flipped for a better comparison. Statistical uncertainties do not exceed 0.02\,cm/s. For more details about calculations, see the text.}
\label{profils_rms}
\end{center}
\end{figure}

The most remarkable fact concerns the larger fluctuations intensity in the $\mathcal{RS}$ cell compared to the $\mathcal{SS}$ one. The average horizontal velocity fluctuations RMS in the $\mathcal{SS}$ cell is $0.43\pm0.01$\,cm/s against $0.56\pm0.01$\,cm/s for the $\mathcal{RS}$ one. It corresponds to a 30$\%$ increase. We observe a similar change for the vertical velocity fluctuations RMS with a 23$\%$ increase (from $0.47\pm0.01$\,cm/s against $0.58\pm0.01$\,cm/s). These fluctuations may have for origin the destabilization of the thermal boundary layers by roughness observed by Salort \emph{et al.} \cite{salort2014} and the pending transition to turbulence of the viscous boundary layer \cite{liot2016}. This destabilization induces probably an increase of plumes emission and/or intensity which leads to a crucial increase of the velocity fluctuations. 

The spatial structure of horizontal velocity fluctuations RMS field for the $\mathcal{RS}$ cell shows a clear bottom-top asymmetry. A zone of large fluctuations appears where the hot jet impacts the top plate. Then these fluctuations spread along this plate. The cold jet spreads also along the hot plate but with a smaller effect. Moreover, this phenomenon is observable for vertical velocity fluctuations RMS too. These zones of large fluctuations are due to the impact of jets on plates, but this asymmetry could be explained by difference in plumes structure or distribution: more intense and/or more numerous plumes starting from the rough plate could be an explanation to this asymmetry. However, given the intensity of the asymmetry, experimental errors cannot be incriminated. Indeed, the RMS velocity fluctuations fields remain symmetric in the $\mathcal{SS}$ cell.

To make these observations more visible, we plot horizontal velocity fluctuations RMS profiles on the figure \ref{profils_rms}. For each cell, they are computed by averaging in a horizontal band of ten-centimeter height starting from the bottom plate and a similar band starting form the top one. Profiles computed in the "bottom" region are horizontally flipped on the graph for a better comparison. We observe quantitatively the rise of fluctuations in presence of roughness. Moreover, the bottom-top asymmetry is very clear and we see fluctuations up to 17\% higher in the profile computed at the top of the cell compared to the one computed at the bottom. It confirms observations made on global fields.

\subsection{Velocity structure functions}

\begin{figure*}
\begin{center}
\includegraphics[width=16cm]{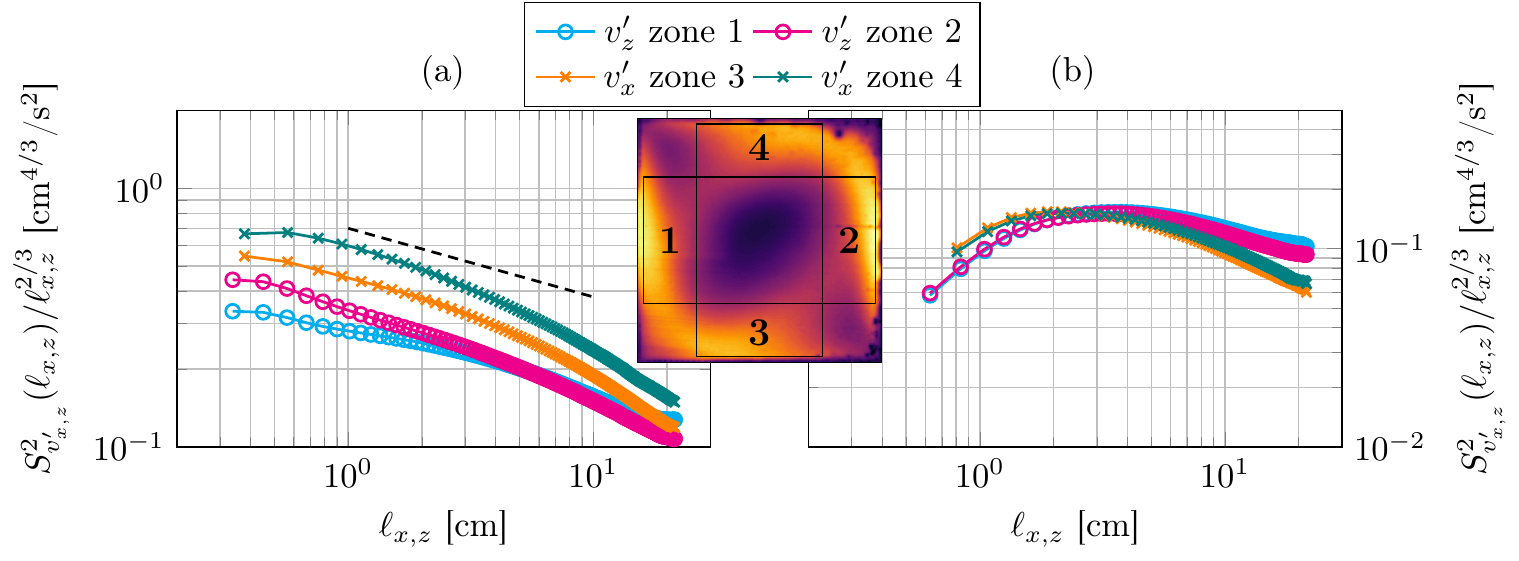}
\caption{Comparison of the velocity fluctuations longitudinal structure functions in zones 1 to 4 (a) in the $\mathcal{RS}$ cell and (b) in the $\mathcal{SS}$ one. Vertical axis are compensated by $\ell_{i}^{2/3}$ where $i=x,z$. The dashed line on figure (a) corresponds to $S^2_{v'_i}(\ell_i)\propto\ell_i^{0.4}$. Inset: zones where statistics are computed; see text for exact boxes dimensions.}
\label{fns_struct_K41}
\end{center}
\end{figure*}

Since we observed an increase of velocity fluctuations in presence of roughness, we can wonder if the turbulence structure is modified. In our case we focus on longitudinal structure functions in specific zones of the flow. We choose to calculate these quantities where both mean velocity has a constant direction and velocity fluctuations are quite homogeneous. Inset of the figure \ref{fns_struct_K41} shows this cutting. Zones 1 and 2 are 21.5\,cm in height and 10\,cm in width and coincide with cold and hot jets respectively. Zones 3 and 4 are 10\,cm in height and 21.5\,cm in width. Each zone starts at 1\,cm from the boundaries. We define longitudinal second-order structure functions of the velocity fluctuations ($v'_{i},~i=x,z$) as :

\begin{equation}
\left \{
\begin{array}{c @{~=~} l}
S^2_{v'_x}(\ell_x) & \langle\left(v'_x\left(x+\ell_x,z,t\right)-v'_x\left(x,z,t\right)\right)^2\rangle_{x,z,t}  , \\
S^2_{v'_z}(\ell_z) & \langle\left(v'_z\left(x,z+\ell_z,t\right)-v'_z\left(x,z,t\right)\right)^2\rangle_{x,z,t}. \\
\end{array}
\right.
\end{equation}

\noindent where $\ell_x$ and $\ell_z$ are the longitudinal spatial increments. $S^2_{v'_x}$ is computed in zones 3 and 4 and $S^2_{v'_z}$ is computed in zones 1 and 2. \textcolor{black}{According to a numerical study from Kaczorowski \emph{et al.} \cite{kaczorowski2014}, the structure function calculation in thermal convection is affected by the spatial resolution. They suggest that this last must be at least similar to the boundary layer typical size (1\,mm in our case). Unfortunately we only reach 3\,mm in resolution with our PIV measurements. Nevertheless, the global trend of the structure functions is not significantly affected by the resolution in this numerical work \cite{kaczorowski2014}.}

Figures \ref{fns_struct_K41}\,(a) and \ref{fns_struct_K41}\,(b) show the longitudinal velocity fluctuations structure functions in the different zones for the $\mathcal{RS}$ and the $\mathcal{SS}$ cells respectively. Structure functions are compensated by $\ell_{i}^{2/3},~i=x,z$ to be compared to the well-known K41 scaling \cite{kolmogorov1941,Monin}:

\begin{equation}
S^2_{v'_{i}} = C_2\left(\epsilon \ell_{i}\right)^{2/3},
\label{eq:fns_struct}
\end{equation}

\noindent where $\epsilon$ is the kinetic energy dissipation rate, $C_2$ the Kolmogorov constant and $i=x,z$. For the $\mathcal{SS}$ situation we observe a very short plateau for every structure functions which is a signature of a short inertial range with K41 behaviour. To assess the turbulence strength we use the Reynolds number based on the Taylor microscale: 

\begin{equation}
R_\lambda = \left(v^{RMS}\right)^2 \sqrt{\frac{15}{\nu\epsilon}}.
\end{equation}

\noindent We estimate $\epsilon$ as \cite{shraiman1990}:

\begin{equation}
\epsilon=\frac{\nu^3}{H^4}RaPr^{-2}(Nu-1),
\label{eq:eps}
\end{equation}

\noindent which leads to $R_\lambda\approx 35$ for the $\mathcal{SS}$ cell and $R_\lambda\approx 60$ for the $\mathcal{RS}$ one. Low $R_\lambda$ makes the inertial range difficult to observe. It explains why the plateaus on compensated $\mathcal{SS}$ velocity fluctuations structure functions are so short. Nevertheless we can discuss the plateau level linked to the prefactor $C_2\epsilon^{2/3}$ (eq. \ref{eq:fns_struct}). The top of the compensated structure functions reach about $0.16\pm0.01$\,cm$^{4/3}$/s$^{2}$. $C_2$ is about constant for $R_\lambda>100$ ($C_2\approx2.1$), but for our Reynolds number we can assess that $C_2\approx1.6$ \cite{sreenivasan1995}. In the considered zones, it leads to a spatial averaged kinetic energy dissipation rate estimated from velocity structure functions $\epsilon_{sf}=3.2\pm0.4\times10^{-2}$\,cm$^2$/s$^{3}$. From eq. \ref{eq:eps} the kinetic energy dissipation rate averaged on the whole cell reaches $\epsilon\approx6.4\times10^{-2}$\,cm$^2$/s$^3$. However the local value of $\epsilon$ is largely inhomogeneous in the flow and depends on the spatial position in the cell, as shown by a \textcolor{black}{numerical study from Kaczorowski \emph{et al.} \cite{kaczorowski2014}. They performed their simulations in a rectangular cell with the same vertical and horizontal aspect ratios as our experiment, for $Pr=4.38$ and $Ra=1\times10^{10}$. They show that the mean kinetic energy dissipation rate in a parallelepipedic sub-volume with dimensions of a quarter of the entire cell in the center of the cell represents only few percents of the global $\epsilon$ computed using eq. \ref{eq:eps}. We want to assess quantitatively the kinetic energy dissipation rate in the zones described in the inset of the figure \ref{fns_struct_K41} which are quite far from the cell center. Consequently we have to use another numerical study }of Kunnen \emph{et al.} \cite{kunnen2008} \textcolor{black}{performed in a cylindrical geometry, even if the LSC global moves due to this geometry could slightly change the kinetic energy dissipation spatial distribution}. Large values of $\epsilon(x,z)$ are observed very close to the boundaries whereas in the rest of the cell $\epsilon(x,z)$ is up to two orders of magnitude lower. In the zones where we compute the structure functions we voluntarily exclude parts of the flow very close to the boundaries. Using results from Kunnen \emph{et al.} \cite{kunnen2008} \textcolor{black}{(performed for $Pr=6.4$ and $Ra=1\times10^9$)} we can estimate that the real averaged $\langle\epsilon(x,z)\rangle_{j}$ in our zones ($j\in\{1,2,3,4\}$) is between 50 and 70\% the one computed from the equation \ref{eq:eps}. Consequently we obtain $\langle\epsilon(x,z)\rangle_{i}=3.8\pm0.6\times10^{-2}$\,cm$^2$/s$^3$. We have a \textcolor{black}{quite} good agreement between $\langle\epsilon(x,z)\rangle_{i}$ corrected by the inhomogeneity and $\epsilon_{sf}$ assessed from $S^2_{v_{i}},~i=x,z$. \textcolor{black}{The small difference could be due to the $Ra$, $Pr$ or geometry difference.}

While the velocity longitudinal structure functions observed in the $\mathcal{SS}$ cell are compatible with the Kolmogorov theory, they \textcolor{black}{differ significantly} in the $\mathcal{RS}$ cell. A lower scaling appears in zones 2 to 4, compatible with $\ell_{i}^{0.4},~i=x,z$. In the zone 1 the scaling seems slightly higher but does not reach the K41 one. In this zone plumes from the top smooth plate are dominant so it is consistent to observe a different scaling from other zones where plumes from the rough plate are dominant -- because they are emitted closely (zone 3) or advected (zones 2 and 4). Moreover we observe that in the zone 4, $S^2_{v_x}$ is larger than in the zone 3. It is consistent with the bottom-top fluctuation asymmetry observed in the cell (figure \ref{RMSvx}). This difference is less visible on $S^2_{v_z}$ because the zones where it is computed do not capture the asymmetry. This dramatic change of the longitudinal velocity structure functions denotes a large change of the turbulence structure. To our knowledge it does not match with theoretical predictions.


\section{Viscous boundary layer structure}

This dramatic change in the turbulence structure can be linked to the evolution of plume intensity and emission from the hot plate that we observed looking at the fluctuation maps (figures \ref{RMSvx} ans \ref{RMSvy}). This change of regime \textcolor{black}{could be} linked to a change of the boundary layer structure after the transition of \textcolor{black}{$Nu$--$Ra$} regime. As pointed in the introduction, some important changes were observed in the thermal boundary layer in the same cell. We previously showed \cite{salort2014} that the thermal boundary layer above the top of obstacles is thinner than in the case of a smooth plate, which can be linked to a heat flux enhancement. A hypothesis of destabilization of the boundary layers for geometric reasons was proposed to build a model to explain the heat flux increase. This hypothesis was recently confirmed by the study of the viscous boundary layer by PIV in a six-time larger proportional cell filled with air built in the \emph{Barrel of Ilmenau} at a similar Rayleigh number \cite{liot2016}. A logarithmic layer, signature of a turbulent boundary layer, was revealed.

In our cell filled with water, it is much more complicated to carry out PIV measurements very close to the roughness. First there are parasite reflections due to particles seeding on the plate. Moreover, intense temperature fluctuations close to the plates lead to large index fluctuations which disturb the visualization. However the logarithmic layer menti\textcolor{black}{o}ned before develops quite far enough from the plate to be observed in the cell. We perform measurements for $Ra=4\times10^{10}$ (see table \ref{table:tableau2}) which is one order of magnitude higher than the transition Rayleigh $Ra_t$. PIV is carried out horizontally centered in the cell (see figure \ref{cell}). We visualize about two obstacles and two notches. To study the shape of the velocity profiles we use the same framework as we previously proposed \cite{liot2016} and commonly employed for logarithmic layers \cite{schlichting2000}. We estimate a friction velocity using \textcolor{black}{the same method as in a previous similar work \cite{liot2016}. It is defined as}
\textcolor{black}{
\begin{equation}
\tau=\rho U^{*2},
\end{equation}
}

\noindent \textcolor{black}{where $\rho$ is the density of the fluid and $\tau$ is the shear stress \cite{schlichting2000}. This last can be linked to the of the Reynolds tensor and the velocity gradient:}

\textcolor{black}{
\begin{equation}
\tau=\rho\langle u'v'\rangle_t + \nu\rho\dfrac{\partial u}{\partial z}.
\end{equation}
}

\noindent where $u'$ and $v'$ represent the velocity fluctuations of horizontal and vertical velocities respectively. \textcolor{black}{In the the described experiment $\tau$ is computed quite away from the plate so we can neglect the velocity gradient and}

\textcolor{black}{
\begin{equation}
U^*\approx \sqrt{\langle u'v'\rangle_t}.
\end{equation}
}

\noindent \textcolor{black}{Finally we estimate $U^*$ using} the maximum value of the Reynolds tensor: 

\begin{equation}
U^*=\text{max}\left(\sqrt{\langle u'v'\rangle_t}\right),
\end{equation}

\noindent  We have $U^*\approx0.35$\,cm/s. Then we define a non-dimensional altitude above the rough plate: 

\begin{equation}
z^+=\frac{zU^*}{\nu}.
\end{equation}

\begin{figure}
\begin{center}
\includegraphics[width=8cm]{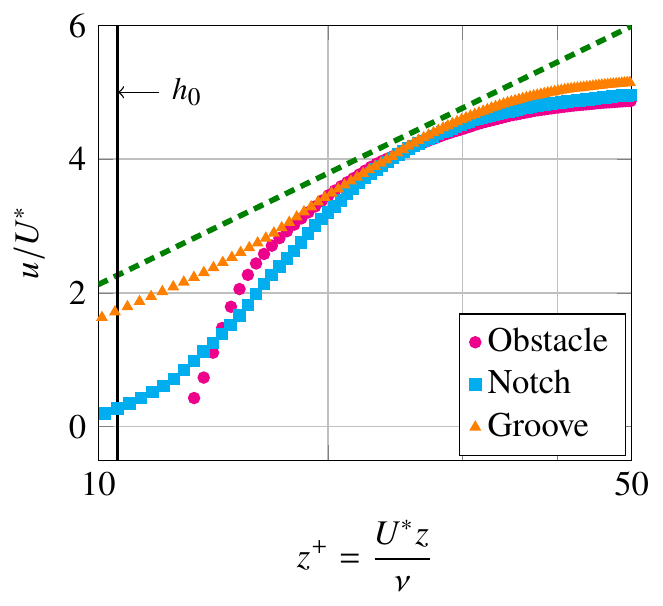}
\caption{Horizontal velocity normalized by $U^*$ function of the normalized altitude $z^+$ for the three locations (described on figure 2). The dashed line represents the equation \ref{eq:log} for $B=-3.4$. The origin for $z$ is taken at the plate in the groove/notch. Consequently the obstacle profile starts at $h_0$. The first points of the profiles are removed because of lack of resolution very close to the plate.}
\label{profils_rough}
\end{center}
\end{figure}

We plot figure \ref{profils_rough} the horizontal velocity normalized by $U^*$ in front of $z^+$ for the three locations (above the obstacle, groove, notch). \textcolor{black}{The resolution of the velocity profiles coupled to a logarithmic scale crush the region of interest. That's why we plot the profiles only between $z^+=10$ and $z^+=50$.} They are compared to the logarithmic profile \cite{schlichting2000}: 

\begin{equation}
\frac{u}{U^*} = 2.40 \ln z^+ +B,
\label{eq:log}
\end{equation}

\noindent with $B=-3.4$. We observe a \textcolor{black}{short} logarithmic layer from $z^+\approx20$ particularly visible in the groove. Moreover the profiles are consistent with the prefactor 2.4. Nevertheless, the estimation of $U^*$ should be considered with care because of the lack of resolution close to the roughness and could affect the adimensioned profiles. For a smooth plate, the expected constant $B$ is 5.84. In order to know if the surface is hydrodynamically rough, we compare the roughness height $h_0$ to the viscous sub-layer $\delta$ that \textcolor{black}{can be estimated} by \cite{tennekes1987}:

\begin{equation}
\delta\approx5\frac{\nu}{U^*}.
\end{equation}

\noindent We have $\delta\approx1.1$\,mm which is smaller than $h_0$. Consequently $B$ depends on $k^+$ defined by: 

\begin{equation}
k^+=\frac{h_0U^*}{\nu} 
\end{equation}

\noindent which reaches about 11 here. The fully rough regime coincides with $k^+\gg 100$ but a transition regime appears for $k^+>5$ \cite{schlichting2000,tennekes1987} which is our case. For sand-shape roughness, the experimental values of $B$ in the transition regime are in the range $[-5,5]$. Our observations are consistent with this assertion. 

We can now assess the thickness of the thermal boundary layer $\delta_{th}$. This one is expected to be thinner than the viscous sub-layer. We adopt the same point of view as our previous study in a rough cell filled with air \cite{liot2016} An analytical and numerical study from Shishkina \emph{et al.} \cite{shishkina2013} has shown that the ratio between the thermal and the viscous ($\delta_v$) boundary layers thickness is highly dependent from the attack angle $\beta$ of the wind on the plate. We extrapolate their results obtained for a laminar boundary layers so the following discussion should be understood in term of order of magnitude. For a laminar boundary layer the ratio $\delta_{th}/\delta_{v}$ ranges from 0.60 ($\beta=\pi/2$) to 1.23 ($\beta=\pi$) for $Pr=4.38$. The flow in the log-layer is turbulent so the attack angle $\beta$ does not remain constant. If we use this study to our viscous sub-layer ($\delta\approx1.1$\,mm), we can only assume that

\begin{equation}
0.66\,mm\leq\delta_{\theta}\leq 1.35\,mm.
\end{equation}

\noindent Using the model developed by Salort \emph{et al.} \cite{salort2014} in the same cell, the expected thermal boundary layer thickness above an obstacle is 0.64\,mm which is consistent with this estimation for high $\beta$.

Finally we find back \textcolor{black}{hints of a} logarithmic layer above roughness, in good agreement with experiments carried out in the air \cite{liot2016}. It is an other \textcolor{black}{clue} that in the case of square-studs roughness a turbulent boundary layer \textcolor{black}{could} develop above the roughness for Rayleigh numbers higher than $Ra_t$. This turbulent boundary layer participates fully to the heat flux increase.


\section{Discussion and conclusion}

We have observed a large increase of the velocity fluctuations in presence of roughness compared to two smooth plates. This increase is visible in the whole cell with the appearance of a bottom-top asymmetry and \textcolor{black}{several indications (e.g. the bottom-top velocity fluctuations asymetry) let think that it could be} attributed to an increase of plumes emission. Moreover, we have confirmed for $Pr=4.4$ \textcolor{black}{a possible} transition to a turbulent boundary layer close to roughness, as already observed in a proportionnal configuration and at a similar $Ra$ for $Pr=0.7$ \cite{liot2016}. Sharp roughness edges are known to be a source of plumes emission as observed at the top of pyramidal roughness \cite{du2000,qiu2005}. But a turbulent boundary layer could boost the plume emission too -- without being discordant with the sharp-edge mechanism. \textcolor{black}{A turbulent boundary layer necessarily implies a logarithmic mean-temperature profile \cite{kraichnan1962,grossmann2012}.} A recent numerical study \cite{vanderpoel2015} for $Pr=1$ and $Ra=5\times10^{10}$ has shown that the plume emission regions on a smooth plate correspond to zones of boundary layer \textcolor{black}{where the mean-temperature profile is logarithmic} whereas \textcolor{black}{the locations of the plate with no plume emission does not reveal such a temperature profiles}. In our case, the wind shear above roughness destabilizes the boundary layer for geometric reasons. Consequently the boundary layer \textcolor{black}{could} transit to turbulence on the whole rough plate so the plume emission by the bottom plate could be globally increased according to numerical observations cited above \cite{vanderpoel2015}. It is consistent with previous background-oriented synthetic Schlieren measurements \cite{salort2014} in the same cell which have shown that the plume emission seems to be homogeneous along the rough plate. Then plumes emitted by the rough plate are advected by the mean wind towards the base of the hot jet then towards the cold plate, that why the bottom-top asymmetry is particularly large close to the impacting region of the hot jet (figure \ref{RMSvx}). 

Furthermore the size of thermal plumes could participate to this elevation of velocity fluctuations. It is usually admitted that their typical size is similar to the thermal boundary layer thickness. Yet the thermal boundary layer is thinner above the roughness than above the top smooth plate \cite{salort2014}. But the pattern formed by the roughness could have an effect on plumes size. A hypothesis could be that plumes are emitted either by top of obstacles or by notches. So they could have a typical size similar to the pattern step (in our case 5\,mm) while the smooth thermal boundary layer is about 1\,mm. An increase of plume size could be at the origin of the scaling change of velocity structure functions observed between the $\mathcal{SS}$ and the $\mathcal{RS}$ situations. This assertion is reinforced by the steeper structure function in the zone 1 (cold jet) where plumes from the smooth plates -- not affected by roughness -- are dominant. Unfortunately we do not have more precise explanation of this observation.



Finally the major results of this study is a dramatic increase of velocity fluctuations in the whole cell in presence of roughness on the bottom plate. It is coupled with a scaling change of the longitudinal velocity structure functions close to the plates and the sidewalls. These observations could be linked to the \textcolor{black}{short} logarithmic layer observed above the roughness. \textcolor{black}{Since pointed out by a previous thermometric study \cite{salort2014}, the thermal flux measurements in the literature show some discrepancy between the results from different roughness geometries (e.g. pyramidal \cite{du2000}, V-shape grooves \cite{roche2001} or square-studs \cite{tisserand2011}). The weight of the two mechanisms observed here (transition to a turbulent boundary layer and plume emission increase) could vary for other square-studs roughness aspect-ratio as it was observed for sinusoidal roughness in recent 2D numerical simulations \cite{toppaladoddi2017}.}



\begin{acknowledgments}
We thank Denis Le Tourneau and Marc Moulin for the manufacture of the cell. This study benefited of fruitful discussions and advices from Bernard Castaing. PIV treatments were made possible with the help of PSMN computing resources.
\end{acknowledgments}

\bibliography{biblio_liot_PRF}

\end{document}